\documentclass[10pt,conference]{IEEEtran}
\IEEEoverridecommandlockouts
\usepackage{cite}

\usepackage{amsmath,amssymb,amsfonts}
\usepackage{algorithmic}
\usepackage{graphicx}
\usepackage{textcomp}
\usepackage{subcaption}
\usepackage{tabularx}
\usepackage{xcolor}
\usepackage{xspace}
\usepackage[inline]{enumitem} 
\usepackage{booktabs}
\usepackage{multirow}
\usepackage{graphicx}
\usepackage{hyperref}
\graphicspath{{img/}}
\newcommand{\cezar}[1]{{\color{blue}\emph{Cezar says: #1}}\xspace}

\begin{document}

\title{\texttt{LabelGit}: A Dataset for Software Repositories Classification using Attributed Dependency Graphs} 

\author{\IEEEauthorblockN{Cezar Sas}
\IEEEauthorblockA{\textit{Bernoulli Institute} \\
\textit{University of Groningen}\\
Groningen, Netherlands \\
c.a.sas@rug.nl}
\and
\IEEEauthorblockN{Andrea Capiluppi}
\IEEEauthorblockA{\textit{Bernoulli Institute}  \\
\textit{University of Groningen} \\
Groningen, Netherlands \\
a.capiluppi@rug.nl}
}

\maketitle

\begin{abstract}
Software repository hosting services contain large amounts of open-source software, with GitHub hosting more than 100 million repositories, from new to established ones. Given this vast amount of projects, there is a pressing need for a search based on the software's content and features.
However, even though GitHub offers various solutions to aid software discovery, most repositories do not have any labels, reducing the utility of search and topic-based analysis.
Moreover, classifying software modules is also getting more importance given the increase in Component-Based Software Development. However, previous work focused on software classification using keyword-based approaches or proxies for the project (e.g., README), which is not always available. In this work, we create a new annotated dataset of GitHub Java projects called {\normalfont \texttt{LabelGit}}. Our dataset uses direct information from the source code, like the dependency graph and source code neural representations from the identifiers. Using this dataset, we hope to aid the development of solutions that do not rely on proxies but use the entire source code to perform classification. 

\end{abstract}


\section{Introduction}
 
Given the overwhelming amount of repositories hosted on services like GitHub, there is a need for a functionality that allows for a search and retrieval based on semantics. As a solution, GitHub proposed a service called \textit{Topics}\footnote{\href{https://github.com/topics}{https://github.com/topics}}, which allows user to annotate projects manually and others to search using these topics; and \textit{Collections}\footnote{\href{https://github.com/collections}{https://github.com/collections}} a curated list of topics where repositories are displayed. However, these solutions are not perfect and have various issues. For example, \textit{Topics} are optional, and GitHub does not suggest or restrict the user in any way. As a result, there are plenty of similar (or identical) topics written with various surface forms, making the search less effective. On the other hand, the \textit{Collections} list is manually curated; therefore, it is not scalable to all topics, reducing the effectiveness of finding repositories, especially those annotated with non-popular topics. 

Recently, there has been an increase in the interest of classification of GitHub repositories.

\cite{vargas2015automatic} uses an approach based on bytecode, and the external dependency of the project, with information from stack overflow to generate a tag cloud. Their dataset is unavailable. 

Sharma et al.~\cite{sharma2017cataloging} uses an combined solution of topic modeling and genetic algorithms called LDA-GA\cite{panichella2013ldaga}. They apply topic modeling on the README files, and optimize the hyperparameters using genetic algorithms. While LDA is an unsupervised solution, humans are needed to label the topics from the identified keywords. They release a list of 10,000 examples annotated by their model (half belonging to 'Others' category), which was evaluated using 400 manually labeled projects.

ClassifyHub~\cite{soll2017classifyhub} uses an ensemble of 8 na\"{i}ve classfiers, each using different features (e.g. file extensions, README, GitHub metadata and more) to perform the classification task. They use the InformatiCup 2017\footnote{\href{https://github.com/informatiCup/informatiCup2017}{https://github.com/informatiCup/informatiCup2017}} dataset, which contains 7 categories.

HiGitClass~\cite{zhang2019HiGitClass} uses  approach for modeling the co-occurence of multimodal signal in a repository (e.g. user, name of repository, tags, README and more). They perform the annotation according to a hierarchical taxonomy that is given as an input with keyword for each leaf node. The release a datset with taxonomies for two domains: a machine learning taxonomy with 1600 examples, and a bioinformatics one with 876 projects.

In \cite{sipio2020naive}, they use the content of the README files and source code, represented using TFIDF, as input to a probabilistic model called Multinomial Na\"{i}ve Bayesian Network to recommend possible topics. They used 120 popular topics from GitHub, and released a dataset of around 10,000 distantly annotated projects in different programming languages.

\cite{izadi2020topic} uses names, descriptions, READMEs, wiki pages, and file names concatenated together as input to BERT\cite{devlin-etal-2019-bert}, a neural language model, that creates a dense representation of the input text. Then, to this representation, they apply a fully connected neural network to predict multiple labels. Their dataset is no currently available and contains 152K GitHub repositories in various languages classified using 228 topics.

However, all these solutions perform software classification by classifying proxies of the projects (e.g., README), which might not always be available. This problem gets more accentuated if we want to perform classification on smaller parts of code like components, which require models trained on source code. 

This work's contributions are a new dataset, called \texttt{LabelGit}\cite{sas_cezar_2021_4459080}, containing annotated GitHub projects for software classification. The classification task is performed given an attributed graph with continuous features representing the source code files. We hope this will act as a catalyst for further research on machine learning for software engineering. Moreover, this can also be used to develop and evaluate new deep learning methods for real-world graphs. 

\begin{figure*}[htb!]
    \centering
    \includegraphics[width=0.95\textwidth]{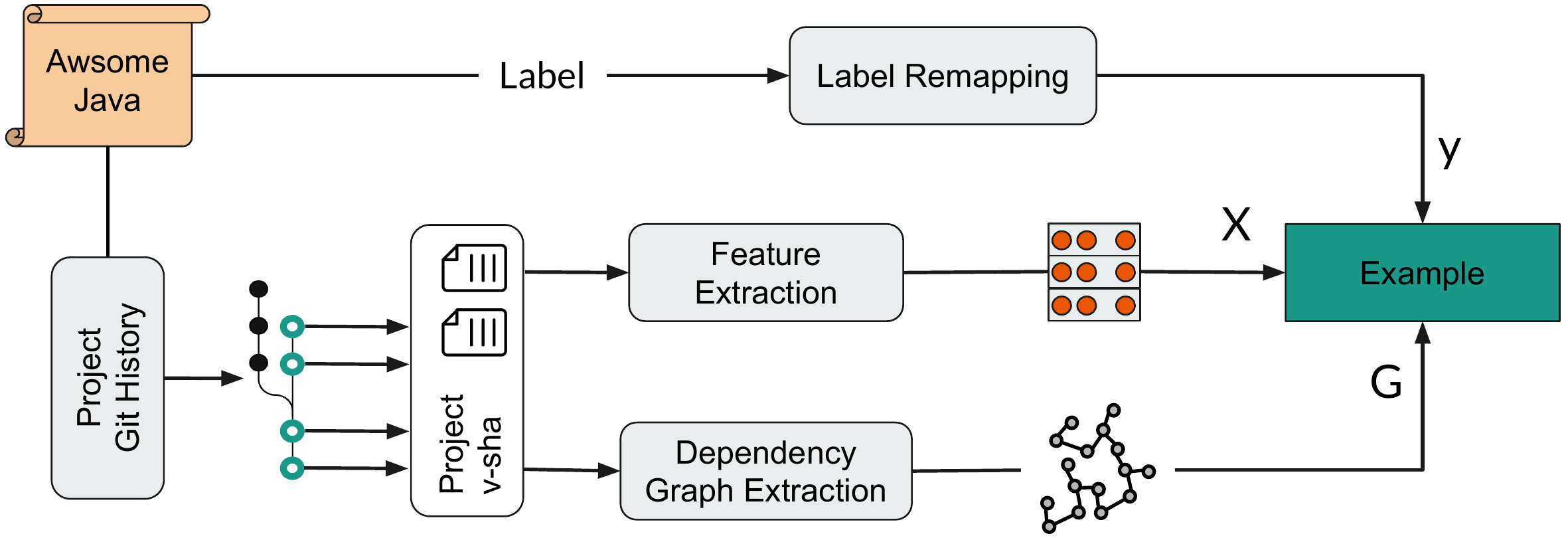}
    \caption{Our dataset creation pipeline. We start with an annotated list from which the annotations are reduced to a smaller set using a manually created mapping. On the other side, we extract the git project history, and for each commit in the \textit{master} branch (green nodes), we run the dependency graph and feature extraction methods. The final example is made by a graph $G$, a feature matrix $X$, and a label $y$.}
    \label{fig:pipeline}
\end{figure*}

\section{Preliminaries}
In the following section, we introduce the concept of creating document embeddings, a dense vector representation of documents used to capture the semantic and syntactic information carried by the specific data source. 

\subsection{Source Code Representations}
The learning of a document's embedding is usually performed using learned representations of words first, then combining those to create a document's embedding. We used two different types of features to represent documents into a continuous space (i.e., its embeddings): the resulting embeddings of the source code files reflect two different interpretations. The first one is \textit{semantic}, and based on natural text information. It is created using a neural language model architecture called \textit{fastText}. The second solution, called code2vec, is trained on source code and learns representations by taking advantage of the \textit{structural} information of source code.

\paragraph{fastText~\cite{bojanowski-etal-2017-enriching}} is a neural language model that creates word embeddings by learning subword information~\cite{bojanowski-etal-2017-enriching}. \textit{fastText} words are split into \textit{n}-grams, and for each it learns an embedding. The final word embedding is an aggregation of the embeddings of all \textit{n}-grams that the word is made of. 
An advantage of using subword information like \textit{n}-grams, is that the smaller information level means that there is no risk of out-of-vocabulary words when applying to more technical domains. 

\paragraph{code2vec\cite{uri2019code2vec}} is a neural embedding model specifically designed for source code. The embeddings are learned using abstract syntax trees paths between variables in a method, and the task is self supervised by predicting the method name. The available pretrained model is learned from Java projects, therefore, the variables representations contain domain specific information.

\section{Datasources}
We created our dataset using two sources, the first is GitHub, while the second is a list of curated GitHub Java projects called \textit{Awesome-Java}. 

\paragraph{GitHub} is a repository hosting service that provides an ecosystem for storing, development, and sharing of software. It is home to many open source communities and commercial software developers worldwide and contains more than 100 million repositories.

\paragraph{Awesome Lists} are curated repositories containing resources that are useful for a particular domain. In our case we use \textit{Awesome-Java}\footnote{\url{https://github.com/akullpp/awesome-java}}, a GiHub project that aggregates several hundreds of curated Java frameworks, libraries and software. It contains 69 categories, and an overall 700 repositories. We removed tutorials, URLs to websites, and projects that failed to be analyzed, and obtained a grand total of 495 projects.

\section{LabelGit}
\texttt{LabelGit}\cite{sas_cezar_2021_4459080} is our new dataset for classification of Java repositories that uses  source code semantics and dependency graphs. It is based on a curated and annotated list of Java projects. 
The pipeline used to create the \texttt{LabelGit} dataset is represented in Figure \ref{fig:pipeline}, and can be divided in the following steps:
\begin{enumerate*}[]
    \item extraction of the dependency graph;
    \item extraction of the features;
    \item label remapping;
    \item data augmentation.
\end{enumerate*}

We made our source code\footnote{\href{https://github.com/SasCezar/LabelGit}{https://github.com/SasCezar/LabelGit}} and dataset\footnote{\href{https://zenodo.org/record/4459080\#.YBPENOhKiF4}{https://zenodo.org/record/4459080\#.YBPENOhKiF4}} publicly available. It contains the annotations in CSV format, the zip file for the graphs, and the two other zip files for the features. The dataset size is around 15 GB compressed, and 52 GB when uncompressed.

\subsection{Dependency Graph}
The first step of our approach is the extraction of the dependency graph for each project in our sample. Using the Arcan~\cite{fontana2017arcan} tool, we obtained the complete set of nodes and the edges describing the dependencies between classes. Edge weights were also extracted to describe the number of uses~\cite{pruijt2017accuracy} of one class by the other. For example, the \textit{antlr4} project contains a dependency between \texttt{ParserATNFactory.java} (from the \texttt{org.antlr.v4.automata} package) and \texttt{LexerGrammar.java} (from the \texttt{org.antlr.v4.tool} package), as the former imports the latter. 

\begin{figure}[htb!]
    \centering
    \includegraphics[width=\columnwidth]{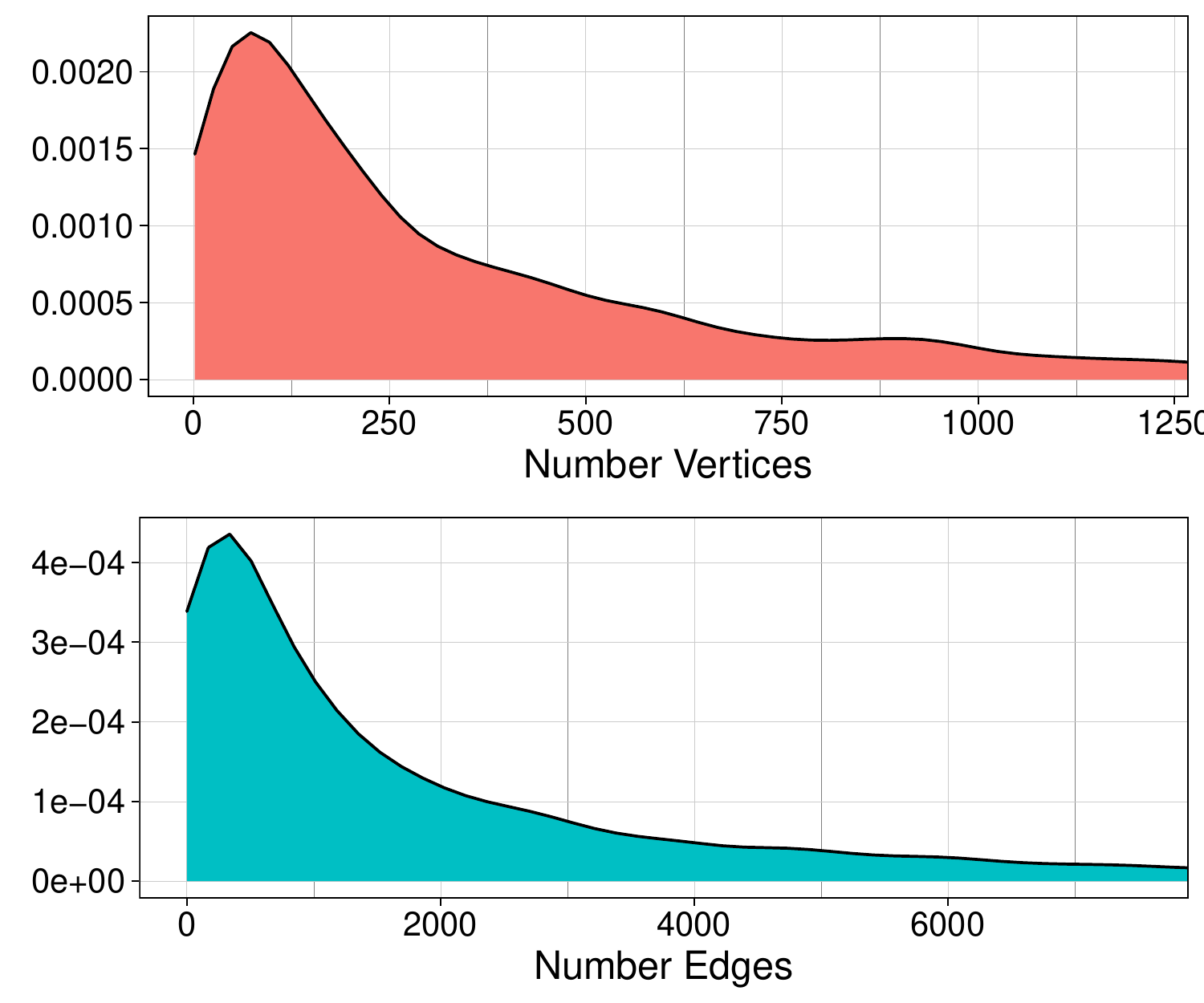}
    \caption{Distribution of the number of vertices (top) and edges (bottom) in the augmented dataset's graphs.}
    \label{fig:graph_distr}
\end{figure}

The distribution of the number of vertices and edges for the graphs in our dataset are shown in Figure \ref{fig:graph_distr}.

The dependency graph is stored using \textit{GraphML} (\texttt{.graphml}), a XML based graph storage file format. The nodes in the graph contain an attribute with the relative path of the file in the project folder, this is used to identify the source file in the feature files and perform alignment.

The \texttt{.graphml} file name allows to easily identify the project and the git commit, and it is structured as follows: 
\begin{center}
    
\textit{[\textit{project name}]-[\textit{commit number}]-[\textit{commit sha}].graphml}
\end{center}

where \textit{[commit number]} is the index of the commit in the version history tree, and \textit{[commit sha]} is the id of the commit.

\subsection{Semantic Features}
The extraction of features representing the documents is performed with two different language models: \textit{fastText} and code2vec. The first one was trained with natural text, the second with source code in various programming languages.

We apply the two embedding methods on the \textit{identifiers} extracted from the source code file. For the extraction of the identifiers contained in the source code, we used the \textit{tree-sitter}\footnote{\href{https://github.com/tree-sitter/tree-sitter}{https://github.com/tree-sitter/tree-sitter}} parser generator tool. It makes easy to get the identifiers, without keywords, from the annotated concrete syntax tree created using a grammar for Java code. We clean the identifiers by separating the camel case strings into words, lower case every token, and remove common Java terms that do not add much semantically (e.g., `\textit{main}', `\textit{println}', etc).

The features are stored in a textual file (\texttt{.vec}), with each file containing the embeddings for a particular project at a specific commit. The project's source code files are described with a row containing the relative path of the file, and the vector of features with each feature separated by a space. 

The filename follows the same structure as before, with the only difference in the extension:
\begin{center}
\textit{[\textit{project name}]-[\textit{commit number}]-[\textit{commit sha}].vec}
\end{center}

\subsection{Label Mapping}
The \textit{Awesome-Java} list contains 69 annotations (or categories), and on average, each category contains some 8 projects, making any classification task very challenging. Moreover, those categories have different granularity levels and can represent very general concepts or very detailed keywords. 

For these reasons, we decided to manually reduce the original categories to a smaller set of 13 (see \textit{Label} column in Table \ref{tab:dataset}). This mapping was performed manually, in a hierarchical fashion, by the authors. 

The initial and final annotated labels are stored as a CSV file (\textit{annotations.csv}). The file has the followings schema:
\begin{itemize}
    \item \textbf{project.name}: name of the project;
    \item \textbf{project.desc}: short description of the project from \textit{Awesome-Java};
    \item \textbf{project.link}: URL to the GitHub repository;
    \item \textbf{category}: \textit{Awesome-Java} annotation;
    \item \textbf{category.desc}: short description of the category from \textit{Awsome-Java};
    \item \textbf{label}: mapping of the \textit{Awsome-Java} category into one of the labels of the smaller set.
\end{itemize}

\subsection{Data Augmentation}
Given the small amount of projects contained in the \textit{Awesome-Java} list, and the complexity of the classification task, we decided to perform a `data augmentation'\cite{shorten2019survey} which has been shown to aid generalization in deep learning models in various domains\cite{ronneberger2015u, perez2017effectiveness}, including software engineering\cite{mi2021augmentation} step. Given our data's nature, we can synthetically generate a large amount of data, while maintaining its original integrity. This is done by making use of the history of a project's repository: instead of taking a single snapshot of it, we take all the snapshots from its \textit{master} branch. Therefore, for each annotated project, we are able to extract multiple different examples that can be used as added `synthetic' projects in the training phase.

This augmentation approach allowed us to increase the initial 495 samples to 11,502, with an average increase of 23 times the amount of samples per project. In Table \ref{tab:dataset} we can see the number of examples before and after the augmentation separated per class.

\begin{table}[htb!]
\begin{center}
    \caption{Distribution of the number of examples before (Projects), and after (Samples) the data augmentation process.}
        \begin{tabular}{lccc}
        \toprule
\textbf{Label}                  & \textbf{Projects} & \textbf{Samples} & \textbf{Increase (x)} \\ \midrule
Introspection          & 32     &     744     & 23 \\
CLI                    & 8      &     142     & 17 \\
Data                   & 49     &     1,088   & 22 \\
Development            & 100    &     2,306   & 23 \\
Graphical              & 11     &     226     & 22 \\
 Miscellaneous         & 59     &     1,729   & 20 \\
Networking             & 25     &     503     & 20 \\
Parser                 & 41     &     935     & 22 \\
Scientific/Engineering & 39     &     915     & 23 \\
Security               & 14     &     249     & 18 \\
Server                 & 37     &     727     & 20 \\
Testing                & 42     &     974     & 23 \\
Web                    & 38     &     964     & 25 \\
\midrule
\textbf{Total} & \textbf{495} & \textbf{11,502} & \textbf{23} \\
\bottomrule
        \end{tabular}
\label{tab:dataset}
\end{center}
\end{table}

To evaluate the overlap resulting from the augmentation approach, we measured the graph difference between two consecutive commits of a project. In order to do so, we take the absolute value of the difference of size (sum of the number of vertices and the number of edges) of the two graphs, and normalize by the maximum size of the considered graphs
\footnote{We chose to evaluate this proxy because computing an exact measure would be extremely demanding, as graph edit distance on large graph requires substantial amounts of memory and time.}. 

Formally, given two graphs representing consecutive analyzed versions of a specific project, $\textbf{G}_i = (V_i, E_i)$, $\textbf{G}_j = (V_j, E_j)$, their difference is:

\begin{equation*}
    diff(\textbf{G}_i, \textbf{G}_j) = \frac{abs((|V_i|+|E_i|) -(|V_j|+|E_j|))}{max(|V_i|+|E_i|, |V_j|+|E_j|)}
    \label{eq:graph_diff}
\end{equation*}

where $|\cdot|$ is the cardinality of the set. 

The average difference between the graphs of two subsequent commits is around 12\%, with a mean interval ($abs(i-j)$) between the analyzed consecutive versions of around 550 commits.

\section{Uses and Future Work}
\paragraph{Uses}
the dataset that we have proposed will aid the development and evaluation of new deep learning models, especially those designed to work on non-euclidean data\cite{Bronstein2017geometricdeeplearning}, such as Graph Neural Networks (GNN)\cite{scarselli2009gnn}. These models will allow to solve tasks such as classification of software projects, and more in general of real-world graph data. Moreover, by performing the classification, these models can also learn better graph representation: this, in turn, can be used for other downstream tasks, such as the identification of architectural smells\cite{pigazzini2019architectual}, and software similarity\cite{capiluppi2020similarity}. Furthermore, the learned representation and the longitudinal nature of our dataset allow us to analyse how software semantics changes with the development and increase in the project's size. Lastly, we can transfer the model's knowledge learned from the coarse project to finer parts of software, like components, to classify them and aid their retrieval for reuse purposes during development.

\paragraph{Future Work}
as future improvements to the dataset, we plan to extract a bottom-up \textit{taxonomy} from the data and provide it with multi-label classification capabilities. This is because a single project might encompass different characteristics and contain components that belong to different categories. We foresee that this process will be based on GitHub's topics and that those will be reduced using hierarchical clustering, as shown in our process above. Moreover, we plan to expand the sample by increasing the number of projects and examples, allowing for a better generalization of models.

\paragraph{Threats to validity}
the main threat to our approach is the use of the curated list for both the gathering of the projects, and as the starting point for creating our taxonomy. This choice can be considered arbitrary; however, we use this as a starting point and something to improve on; moreover, the development of various approaches will not be affected by the taxonomy. 

Another threat is the data augmentation technique adopted as it might not allow for good generalization as they are still only 495 projects; however, this is reduced by the large difference between the graphs and the fact that various studies showed how data augmentation improves generalization in low data domains.

\section{Conclusions}
In this work, we proposed \texttt{LabelGit}, a new dataset for software classification of Java repositories. Compared to previous solutions, ours is designed to be a direct representation of the source code by creating an attributed dependency graph, where each node is represented with semantic features. 
We expect that \texttt{LabelGit} will increase the interest in developing better solutions for software classification, and more in general, for real-world attributed graph classification.

\section*{Acknowledgements}
We would like to thank the Center for Information Technology of the University of Groningen for their support and for providing access to the Peregrine high performance computing cluster.

\bibliography{references}
\end{document}